\documentclass[preprint]{ptephy_v1}

\preprintnumber{RIKEN-iTHEMS-Report-24, KYUSHU-HET-303} 

\usepackage{graphicx}

\usepackage{amsmath} 
\usepackage{amsthm} 
\usepackage{url} 
\usepackage{stmaryrd}
\usepackage{tikz}
\usepackage{physics}
\usepackage{bm}
\usepackage{booktabs}
\usepackage{subcaption}

\numberwithin{equation}{section}

\allowdisplaybreaks

\begin{document}

\title{Winding number on 3D lattice}


\author[1]{Okuto Morikawa}
\affil[1]{Interdisciplinary Theoretical and Mathematical Sciences Program
(iTHEMS), RIKEN, Wako 351-0198, Japan}

\author[2]{Hiroshi Suzuki}
\affil[2]{Department of Physics, Kyushu University, 744 Motooka, Nishi-ku,
Fukuoka 819-0395, Japan}





\begin{abstract}%
We propose a simple numerical method which computes an approximate value of the
winding number of a mapping from 3D torus~$T^3$ to the unitary group~$U(N)$,
when $T^3$ is approximated by discrete lattice points. Our method consists of
a ``tree-level improved'' discretization of the winding number and the
gradient flow associated with an ``over-improved'' lattice action. By employing
a one-parameter family of mappings from $T^3$ to $SU(2)$ with known winding
numbers, we demonstrate that the method works quite well even for coarse
lattices, reproducing integer winding numbers in a good accuracy. Our method
can trivially be generalized to the case of higher-dimensional tori. 
\end{abstract}

\subjectindex{B31,B34,B38}

\maketitle

\section{Introduction}
\label{sec:1}
One of the topological numbers, the winding number of a differential mapping
from a 3D closed manifold~$\mathcal{M}$ to the unitary group~$U(N)$,
\begin{equation}
   W_3:=\frac{1}{24\pi^2}\int_{\mathcal{M}}\tr(g^{-1}\mathrm{d}g)^3\in\mathbb{Z},
   \qquad g\in U(N),
\label{eq:(1.1)}
\end{equation}
appears in various branches of physics, ranging from particle physics
(see, for example, Refs.~\cite{Friedan:1982nk,Luscher:1981zq,Seiberg:1984id})
to condensed matter physics (see Ref.~\cite{Schnyder} and references cited
therein). In some occasions, one wants to (approximately) determine the winding
number only given the mapping only discrete points of~$\mathcal{M}$. This
problem, which is closely related to a lattice formulation of the 3D
non-Abelian Chern--Simons theory, is revisited
recently~\cite{Shiozaki:2024yrm,Hamano,Chen:2024ddr,Zhang:2024cjb}.\footnote{%
The winding number is related to the second Chern number on a 4D discrete
lattice. For the computation of the first Chern number on discrete lattice,
there exists a very effective computational
method~\cite{Luscher:1998du,Fukui:2005wr}.} In this paper, we propose a
numerical method which approximately computes the winding number~$W_3$.
Explicitly, we consider the situation in which a mapping from 3D torus~$T^3$ is
given approximately on discrete cubic lattice points. By employing a
one-parameter family of mappings with known winding numbers, we first
demonstrate that a simple discretization of~Eq.~\eqref{eq:(1.1)} with a
``tree-level improvement'' (that eliminates the leading lattice discretization
error) works quite well, approximately reproducing non-trivial integer winding
numbers. Then, as a possible treatment of data given only on a coarse lattice,
we formulate a smearing based on the gradient flow~\cite{Luscher:2010iy}. We
find numerically that, as recently studied in the context of lattice gauge
theory~\cite{Tanizaki:2024zsu},\footnote{This study is closely related to
studies in~Refs.~\cite{GarciaPerez:1993lic,deForcrand:1995bq}.} the gradient
flow associated with an ``over-improved'' lattice action provides a strong
stabilization of the lattice winding number along the gradient flow. A
combination of the discretized winding number and an appropriate gradient flow
thus provides a simple and versatile method to approximately compute the
winding number; we note that this method requires neither the diagonalization
of~$g(x)$ nor the interpolation of~$g(x)$ on the lattice.

\section{Discretization of~$W_3$}
\label{sec:2}
We assume that the group elements $g(x)\in U(N)$ are residing on lattice
points~$x$ (i.e., sites) of the periodic cubic
lattice~$\Lambda_L=(\mathbb{Z}/L\mathbb{Z})^3$, $x\in\Lambda_L$. The Lorentz
indices are labeled by~$\mu$, $\nu$, \dots\ and the unit vector in the
$\mu$~direction is denoted by~$\Hat{\mu}$. We discretize $W_3$
in~Eq.~\eqref{eq:(1.1)} by substituting the derivative by\footnote{Here, the
lattice spacing~$a$ is set to be unity.}
\begin{equation}
  g(x)^{-1}\partial_\mu g(x)\to
   \frac{1}{4}H(x,\mu),
\label{eq:(2.1)}
\end{equation}
where
\begin{align}
   H(x,\mu)&:=
   g(x)^{-1}
   \Bigl\{
   g(x+\Hat{\mu})-g(x-\Hat{\mu})
\notag\\
   &\qquad\qquad\qquad{}
   -\frac{\eta}{6}
   \left[g(x+2\Hat{\mu})-2g(x+\Hat{\mu})+2g(x-\Hat{\mu})-g(x-2\Hat{\mu})
   \right]\Bigr\}
   -\text{H.c.}
\label{eq:(2.2)}
\end{align}
In this expression, the term proportional to the parameter~$\eta$ is
introduced to control the leading discretization error of the order
of~$a^2g^{-1}\partial_\mu^3 g$ ($a$ being the lattice spacing) in the classical
continuum limit. The choice $\eta=1$ eliminates the leading discretization
error but we will later utilize this freedom of~$\eta$ to realize a gradient
flow that stabilizes the lattice winding number along the flow. We then define
a lattice counterpart of~$W_3$~\eqref{eq:(1.1)} by
\begin{equation}
   W_3^{\text{lat}}:=\frac{1}{4^3\cdot24\pi^2}
   \sum_{x\in\Lambda_L}\sum_{\mu,\nu,\rho}\epsilon_{\mu\nu\rho}
   \tr\left[H(x,\mu)H(x,\nu)H(x,\rho)\right].
\label{eq:(2.3)}
\end{equation}

We examine the validity of this very simple prescription by using a
one-parameter family of mappings with known winding numbers. It is the
following mapping from~$T^3$ to~$SU(2)$:\footnote{This is basically identical
to one of mappings considered in~Refs.~\cite{Shiozaki:2024yrm,Hamano}.}
\begin{equation}
   g(\theta):=\Bar{\xi}_0\bm{1}+i\sum_{\mu=1}^3\Bar{\xi}_\mu\sigma_\mu,\qquad
   \Bar{\xi}_A:=\frac{\xi_A}{\sqrt{\sum_{A=0}^3\xi_A^2}},
\label{eq:(2.4)}
\end{equation}
where $\sigma_\mu$ are Pauli matrices and
\begin{equation}
   \xi_0:=m+\sum_{\mu=1}^3(\cos\theta_\mu-1),\qquad
   \xi_\mu:=\sin\theta_\mu,\qquad-\pi\leq\theta_\mu\leq\pi.
\label{eq:(2.5)}
\end{equation}
Since $g(\theta)\to-g(\theta)$ under $\theta\to\theta+\pi$ and~$m\to6-m$,
$W_3$~\eqref{eq:(1.1)} for this mapping is invariant under~$m\to6-m$. This
mapping appears, for instance, in the computation of the Chern--Simons
term~\cite{So:1985wv,Coste:1989wf} and the axial anomaly~\cite{Fujiwara:2002xh}
in lattice gauge theory; there, the parameter~$m$ is related to the number of
species doublers in a lattice Dirac operator. It can be analytically shown that
\begin{equation}
   W_3=\begin{cases}
   0&\text{for $m<0$ and $6<m$},\\
   1&\text{for $0<m<2$ and $4<m<6$},\\
   -2&\text{for $2<m<4$}.\\
   \end{cases}
\label{eq:(2.6)}
\end{equation}

Now, Tables~\ref{table:1} and~\ref{table:2} show
$W_3^{\text{lat}}$~\eqref{eq:(2.3)} with $\eta=0$ (i.e., no improvement)
and~$\eta=1$ (i.e., tree-level improvement) in~Eq.~\eqref{eq:(2.2)},
respectively, for various lattice sizes (i.e., the number of 1D discretization
points)~$L$.
\begin{table}[htbp]
\caption{The lattice winding numbers~$W_3^{\text{lat}}$~\eqref{eq:(2.3)} for the
mapping~\eqref{eq:(2.4)} with $\eta=0$ in~Eq.~\eqref{eq:(2.2)} (i.e., no
improvement). $L$ is the lattice size. These should be compared with the exact
values in~Eq.~\eqref{eq:(2.6)}.}
\label{table:1}
\begin{center}
\begin{tabular}{clllll}
\toprule
$m$&$L=10$&$L=20$&$L=30$&$L=40$&$L=50$\\
\midrule
$-1$&$2.46937\times10^{-3}$&$1.01207\times10^{-3}$&$4.88982\times10^{-4}$&$2.83336\times10^{-4}$&$1.83861\times10^{-4}$\\
$1$&$7.49024\times10^{-1}$&$9.28484\times10^{-1}$&$9.67423\times10^{-1}$&$9.81515\times10^{-1}$&$9.88121\times10^{-1}$\\
$3$&$-1.51316$&$-1.86086$&$-1.93661$&$-1.96402$&$-1.97688$\\
\bottomrule
\end{tabular}
\end{center}
\end{table}
\begin{table}[htbp]
\caption{Same as Table~\ref{table:1} but computed with $\eta=1$
in~Eq.~\eqref{eq:(2.2)}; this value of~$\eta$ eliminates the leading
discretization error in~$W_3^{\text{lat}}$ (i.e., tree-level improvement).}
\label{table:2}
\begin{center}
\begin{tabular}{clllll}
\toprule
$m$&$L=10$&$L=20$&$L=30$&$L=40$&$L=50$\\
\midrule
$-1$&$1.75038\times10^{-3}$&$2.49312\times10^{-4}$&$5.64141\times10^{-5}$&$1.87146\times10^{-5}$&$7.83516\times10^{-6}$\\
$1$&$9.45947\times10^{-1}$&$9.95024\times10^{-1}$&$9.98940\times10^{-1}$&$9.99655\times10^{-1}$&$9.99857\times10^{-1}$\\
$3$&$-1.89778$&$-1.99036$&$-1.99794$&$-1.99933$&$-1.99972$\\
\bottomrule
\end{tabular}
\end{center}
\end{table}

We observe that the winding numbers are basically well approximately reproduced
even with a simple discretization as in~Eqs.~\eqref{eq:(2.2)}
and~\eqref{eq:(2.3)}; this was somewhat surprising to us. Moreover, we
observe that the tree-level improvement~$\eta=1$, which removes the leading
lattice discretization error of~$O(a^2)$ in~$W_3^{\text{lat}}$ improves the
situation drastically; practically, $L\simeq20$ would be sufficient to extract
the integer winding number reliably.

In the above example, the continuous form of the mapping is a priori known and
one can approximate the mapping in an accuracy as much as one wishes. In
practical problems, however, the number of lattice points could be limited due
to various reasons. For such a mapping in a coarse lattice or a noisy form of
mappings, in the next section, we propose a smearing of the configuration by the
gradient flow.

\section{Gradient flow of~$g(x)$}
\label{sec:3}
Going back to the continuum theory, we consider a one-parameter evolution
of the group valued field~$g(x)\in U(N)$ along a fictitious time~$t\geq0$. We
postulate that the evolution is given by
\begin{equation}
   \partial_t g(t,x)
   =g(t,x)\partial_\mu\left[g(t,x)^{-1}\partial_\mu g(t,x)\right],\qquad
   g(t=0,x)=g(x).
\label{eq:(3.1)}
\end{equation}
This ``flow equation'' is actually the gradient flow associated with the
action~$S$, i.e.,
\begin{equation}
   \partial_t g(t,x)
   =-g(t,x)\frac{\delta S}{\delta\eta^a(t,x)}T^a,
\label{eq:(3.2)}
\end{equation}
where we have parametrized the variation of~$g(t,x)$ as
$\delta g(t,x):=g(t,x)\eta(t,x)$
and~$\eta(t,x)=\eta^a(t,x)T^a$ with $U(N)$ generators~$T^a$.\footnote{We adopt
a convention that $T^a$ are anti-Hermitian and normalized
as~$\tr(T^aT^b)=-(1/2)\delta^{ab}$.} The action~$S$ is given by
\begin{equation}
   S:=-\int_{T^3}d^3x\,\tr
   \left[g^{-1}(t,x)\partial_\mu g(t,x)\cdot g^{-1}(t,x)\partial_\mu g(t,x)\right]
   \geq0.
\label{eq:(3.3)}
\end{equation}
Then, noting the relation $\delta(g^{-1}\partial_\mu g)=%
\partial_\mu(g^{-1}\delta g)+[g^{-1}\partial_\mu g,g^{-1}\delta g]$, it is
straightforward to see that Eq.~\eqref{eq:(3.2)} gives the flow
equation~\eqref{eq:(3.1)}. Also, from
$\partial_t(g^{-1}\partial_\mu g)=\partial_\mu\varphi+[g^{-1}\partial_\mu g,\varphi]$, where $\varphi:=\partial_\nu(g^{-1}\partial_\nu g)$, we see that
$\partial_tS\leq0$, the action monotonically decreases along the flow. The
winding number $W_3$~\eqref{eq:(1.1)}, being an integer, is constant along the
flow. Therefore, in the continuum theory, the flow is confined within a
topological sector specified by~$W_3$ and it comes to a stop at a local minimum
of the action (i.e., a classical solution of~$S$). In lattice theory, we want
to approximately realize this property of the gradient flow in continuum.

The gradient flow equation~\eqref{eq:(3.2)} is naturally transcribed to the
lattice theory as
\begin{equation}
   \partial_t g(t,x)
   =-g(t,x)\partial_x^aS^{\text{lat}}T^a,
\label{eq:(3.4)}
\end{equation}
where the Lie-algebra derivative is defined by
\begin{equation}
   \partial_x^a f(g)=\left.\frac{d}{ds}f(ge^{sX})\right|_{s=0},\qquad
   X(y)=\begin{cases}
   T^a&\text{if $y=x$},\\
   0&\text{otherwise}.\\
   \end{cases}
\label{eq:(3.5)}
\end{equation}
From these relations, we have
\begin{equation}
   \partial_tg(t,x)
   =-\sum_{x\in\Lambda_L}\partial_x^aS^{\text{lat}}\partial_x^a S^{\text{lat}}
\label{eq:(3.6)}
\end{equation}
and, assuming that the lattice action~$S^{\text{lat}}$ is real, $S^{\text{lat}}$
monotonically decreases along the flow. We set the form of the lattice action
as
\begin{align}
   S^{\text{lat}}&:=-\frac{1}{16}\sum_{x\in\Lambda_L}\sum_\mu\tr H(x,\mu)^2\geq0,
\label{eq:(3.7)}
\end{align}
where $H(x,\mu)$ is given by~Eq.~\eqref{eq:(2.2)} and here it is understood
that the substitution~$g(x)\to g(t,x)$ is made.

The gradient flow equation~\eqref{eq:(3.4)} can be numerically solved by the
Runge--Kutta method described in~Appendix~C
of~Ref.~\cite{Luscher:2010iy}.\footnote{Our numerical calculation employs
\texttt{Gaugefields.jl} in the JuliaQCD package~\cite{Nagai:2024yaf}. The
actual code set can be found
in~\url{https://github.com/o-morikawa/Gaugefields.jl}; the numerical data for
this paper is stored in \url{https://github.com/o-morikawa/Wind3D}}
In the notation of~Ref.~\cite{Luscher:2010iy}, the Lie-algebra valued
function~$Z(t,x)$ in
\begin{equation}
   \partial_t g(t,x)=Z(t,x)g(t,x)
\label{eq:(3.8)}
\end{equation}
is given by (omitting the argument~$t$)
\begin{align}
   &Z(x)
\notag\\
   &=-g(x)\partial_x^aST^ag(x)^{-1}
\notag\\
   &=-\frac{1}{16}
   \sum_\mu
   \Bigl(
   -\Bigl\{
   g(x+\Hat{\mu})-g(x-\Hat{\mu})
\notag\\
   &\qquad\qquad\qquad\qquad{}
   -\frac{\eta}{6}
   \left[
   g(x+2\Hat{\mu})-2g(x+\Hat{\mu})+2g(x-\Hat{\mu})-g(x-2\Hat{\mu})
   \right]
   \Bigr\}
   H(x,\mu)g(x)^{-1}
\notag\\
   &\qquad\qquad{}
   +g(x)
   \Bigl\{
   H(x-\Hat{\mu},\mu)g(x-\Hat{\mu})^{-1}
   -H(x+\Hat{\mu},\mu)g(x+\Hat{\mu})^{-1}
\notag\\
   &\qquad\qquad\qquad\qquad{}
   -\frac{\eta}{6}
   \bigl[
   H(x-2\Hat{\mu},\mu)g(x-2\Hat{\mu})^{-1}
   -2H(x-\Hat{\mu},\mu)g(x-\Hat{\mu})^{-1}
\notag\\
   &\qquad\qquad\qquad\qquad\qquad\qquad{}
   +2H(x+\Hat{\mu},\mu)g(x+\Hat{\mu})^{-1}
   -H(x+2\Hat{\mu},\mu)g(x+2\Hat{\mu})^{-1}
   \bigr]
   \Bigr\}
   \Bigr)
\notag\\
   &\qquad{}
   -\text{H.c.}
\label{eq:(3.9)}
\end{align}

When $g\in SU(N)$, we may replace $Z(x)$ by its traceless part:
\begin{equation}
   Z(x)\to Z(x)-\frac{1}{N}\tr\left[Z(x)\right]\bm{1}.
\label{eq:(3.10)}
\end{equation}

\begin{figure}[htbp]
\centering
\begin{subfigure}{0.45\columnwidth}
\centering
\includegraphics[width=\columnwidth]{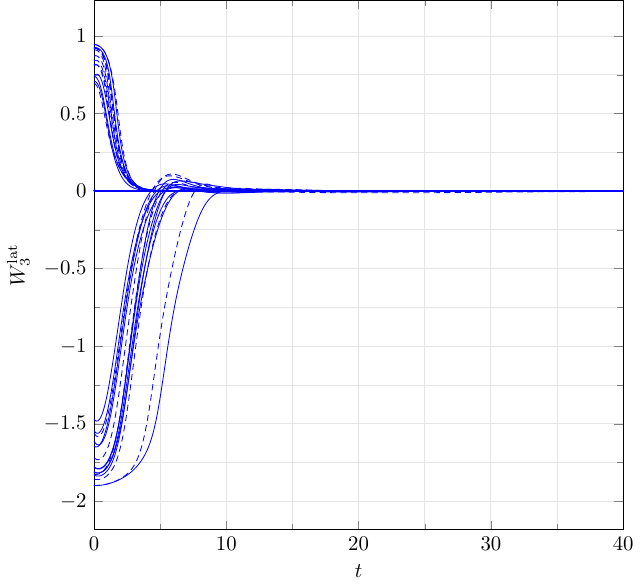}
\caption{$\eta=0$}
\label{fig:1a}
\end{subfigure}
\hspace{6mm}
\begin{subfigure}{0.45\columnwidth}
\centering
\includegraphics[width=\columnwidth]{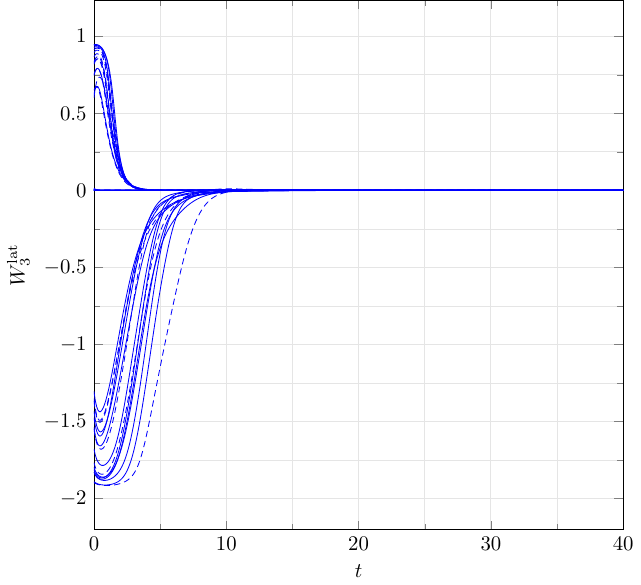}
\caption{$\eta=1$}
\label{fig:1b}
\end{subfigure}
\\[8mm]
\begin{subfigure}{0.45\columnwidth}
\centering
\includegraphics[width=\columnwidth]{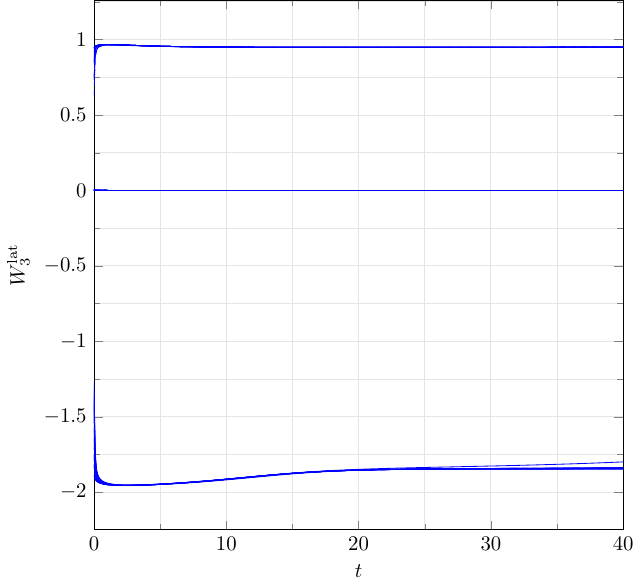}
\caption{$\eta=-10$}
\label{fig:1c}
\end{subfigure}
\hspace{6mm}
\begin{subfigure}{0.45\columnwidth}
\centering
\includegraphics[width=\columnwidth]{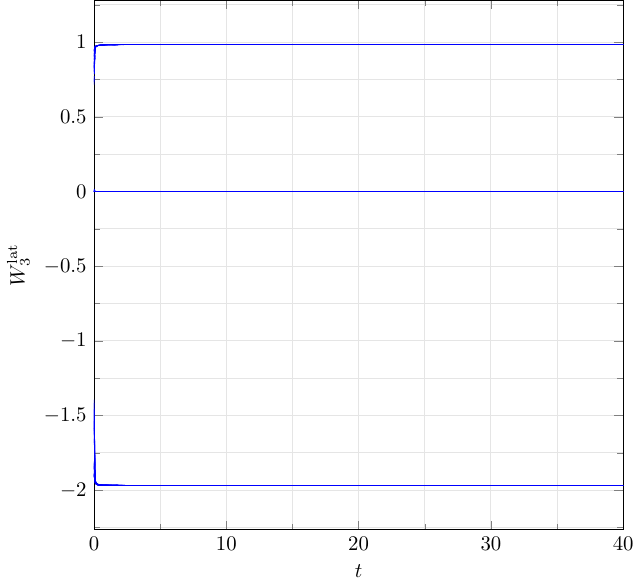}
\caption{$\eta=-20$}
\label{fig:1d}
\end{subfigure}
\caption{$W_3^{\text{lat}}$~\eqref{eq:(2.3)} as the function of the flow time~$t$.
The flow is defined by~Eq.~\eqref{eq:(3.8)} with various values of~$\eta$,
$\eta=0$ (no improvement), $\eta=1$ (tree-level improvement), $\eta=-10$,
and~$\eta=-20$ (``over-improved''), respectively. We set the initial
configuration $g(t=0,x)$ for the flow by Eq.~\eqref{eq:(2.4)} with
$\xi_A(\theta)\to\xi_A(\theta)+\varepsilon r_A(\theta)$, where $\varepsilon$
and $r_A(\theta)$ are uniform random numbers, $\varepsilon\in[0,1]$ and
$r_A(\theta)\in[-1/2,1/2]$. We take 15 random initial configurations for each of
$m=-1$, $m=1$, and~$m=3$; these parameters corresponds to the winding numbers
$W_3=0$, $W_3=1$, and~$W_3=-2$, respectively. The lattice size is~$L=10$.}
\label{fig:1}
\end{figure}

\begin{figure}[htbp]
\centering
\begin{subfigure}{0.45\columnwidth}
\centering
\includegraphics[width=\columnwidth]{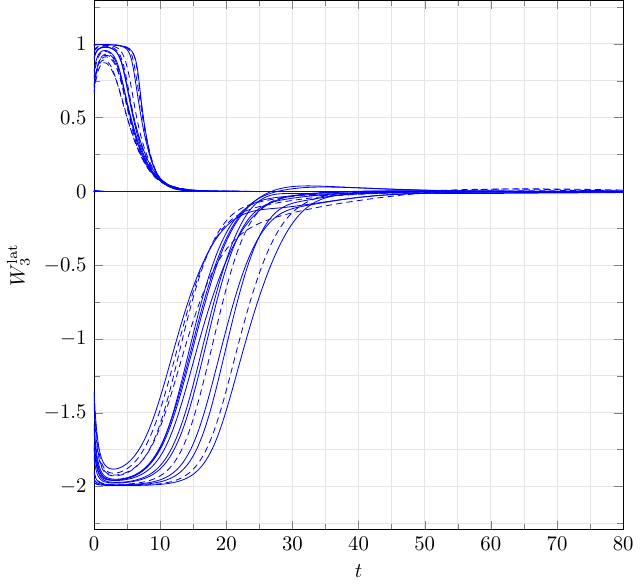}
\caption{$\eta=0$}
\label{fig:2a}
\end{subfigure}
\hspace{6mm}
\begin{subfigure}{0.45\columnwidth}
\centering
\includegraphics[width=\columnwidth]{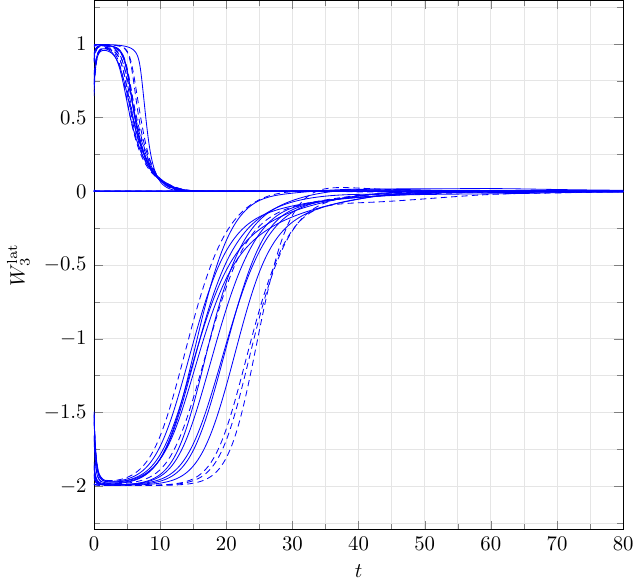}
\caption{$\eta=1$}
\label{fig:2b}
\end{subfigure}
\\[8mm]
\begin{subfigure}{0.45\columnwidth}
\centering
\includegraphics[width=\columnwidth]{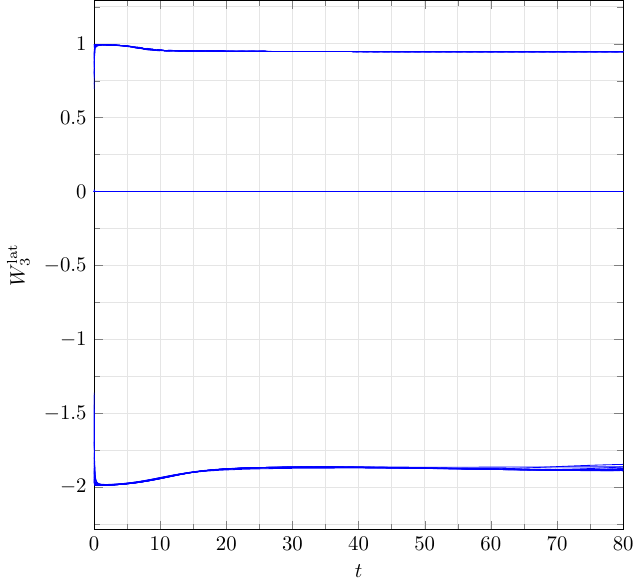}
\caption{$\eta=-10$}
\label{fig:2c}
\end{subfigure}
\hspace{6mm}
\begin{subfigure}{0.45\columnwidth}
\centering
\includegraphics[width=\columnwidth]{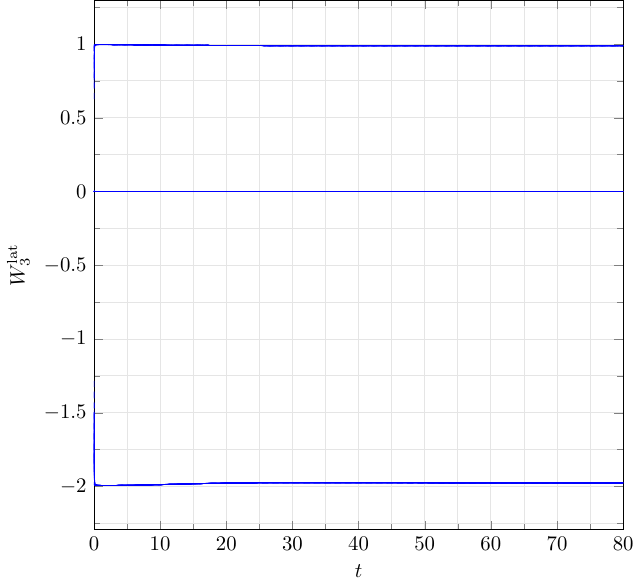}
\caption{$\eta=-20$}
\label{fig:2d}
\end{subfigure}
\caption{Same as~Fig.~\ref{fig:1} but the lattice size is~$L=20$.}
\label{fig:2}
\end{figure}

The gradient flow acts as a smearing on the lattice field. After the flow with
a certain appropriate flow time~$t$, we may employ the discretized formula
for the winding number, Eq.~\eqref{eq:(2.3)}. In Figs.~\ref{fig:1}
and~\ref{fig:2}, we plot $W_3^{\text{lat}}$~\eqref{eq:(2.3)} with~$\eta=1$ (i.e.,
tree-level improved) as the function of the flow time~$t$.\footnote{Since the
diffusion length of the flow equation~\eqref{eq:(3.4)} is~$\sim\sqrt{8t}$, the
flow time larger than $t\sim L^2/32$ would be regarded as an over-smearing.}
The flow is on the other hand defined by~Eqs.~\eqref{eq:(3.8)}
and~\eqref{eq:(3.9)} with various values of~$\eta$, $\eta=0$ (no improvement),
$\eta=1$ (tree-level improvement), $\eta=-10$, and~$\eta=-20$
(``over-improved'' ones), respectively. We set the initial configuration
$g(t=0,x)$ for the flow as Eq.~\eqref{eq:(2.4)} with
$\xi_A(\theta)\to\xi_A(\theta)+\varepsilon r_A(\theta)$, where $\varepsilon$
and~$r_A(\theta)$ are uniform random numbers, $\varepsilon\in[0,1]$ and
$r_A(\theta)\in[-1/2,1/2]$. We take 15 random initial configurations for each
of $m=-1$, $m=1$, and~$m=3$; these parameters correspond to the winding
numbers $W_3=0$, $W_3=1$, and~$W_3=-2$, respectively. The lattice sizes are
$L=10$ in~Fig.~\ref{fig:1}, and $L=20$ in~Fig.~\ref{fig:2}. We observe that the
behaviors drastically change depending on the value of~$\eta$ in the flow; for
$\eta=0$ and~$\eta=1$, the flow acts to eliminate non-trivial winding numbers,
while the over-improved cases ($\eta=-10$ and~$\eta=-20$), the flow stabilizes
the winding number around the desired integer values.\footnote{%
For example, the average of 15 lines in~Fig.~\ref{fig:1d} ($L=10$
and~$\eta=-20$) corresponding to $m=3$ at the flow time $t=2$ gives
$W_3^{\text{lat}}=-1.96797(4)$.} For a usage of the gradient flow associated with
an over-improved lattice action for the winding number stabilization, we got a
hint from a recent study~\cite{Tanizaki:2024zsu} in lattice gauge theory.

\begin{figure}[htbp]
\centering
\begin{subfigure}{0.4\columnwidth}
\centering
\includegraphics[width=\columnwidth]{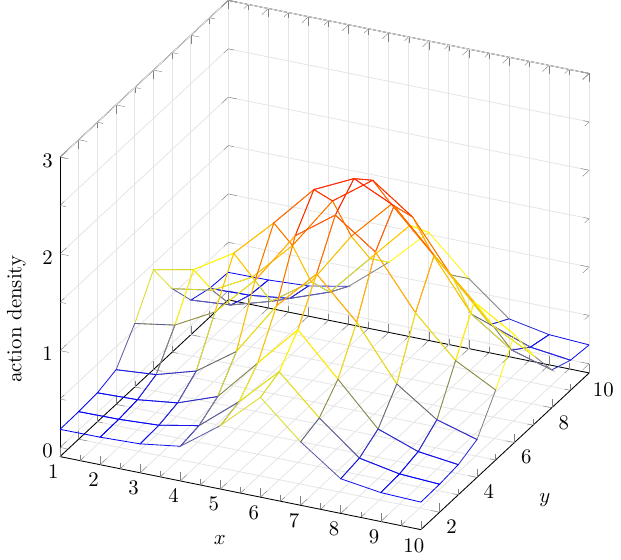}
\caption{$t=0$}
\label{}
\end{subfigure}
\hspace{9mm}
\begin{subfigure}{0.4\columnwidth}
\centering
\includegraphics[width=\columnwidth]{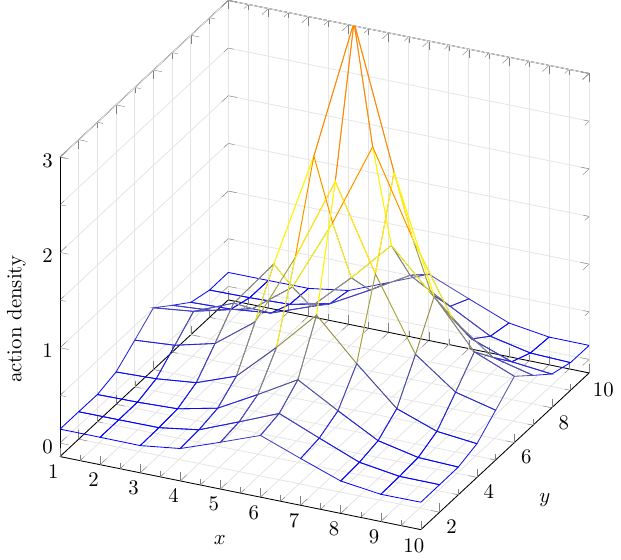}
\caption{$t=1$}
\label{}
\end{subfigure}
\begin{subfigure}{0.4\columnwidth}
\centering
\includegraphics[width=\columnwidth]{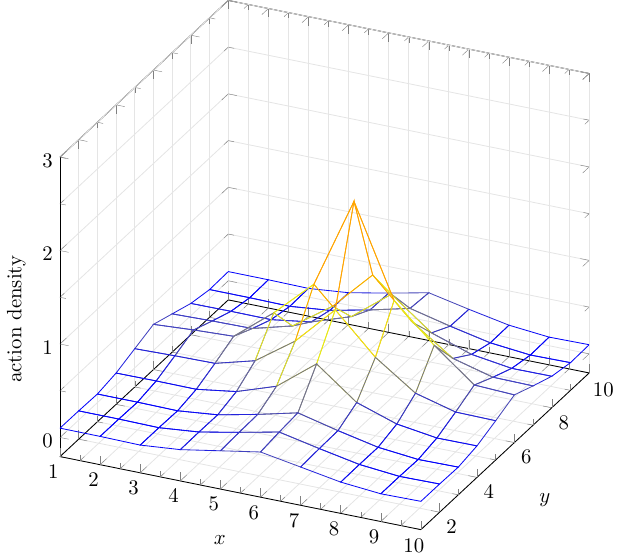}
\caption{$t=2$}
\label{}
\end{subfigure}
\hspace{9mm}
\begin{subfigure}{0.4\columnwidth}
\centering
\includegraphics[width=\columnwidth]{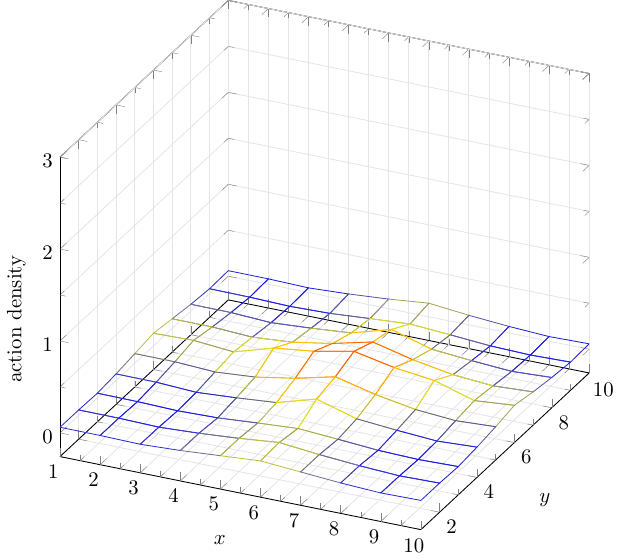}
\caption{$t=3$}
\label{}
\end{subfigure}
\begin{subfigure}{0.4\columnwidth}
\centering
\includegraphics[width=\columnwidth]{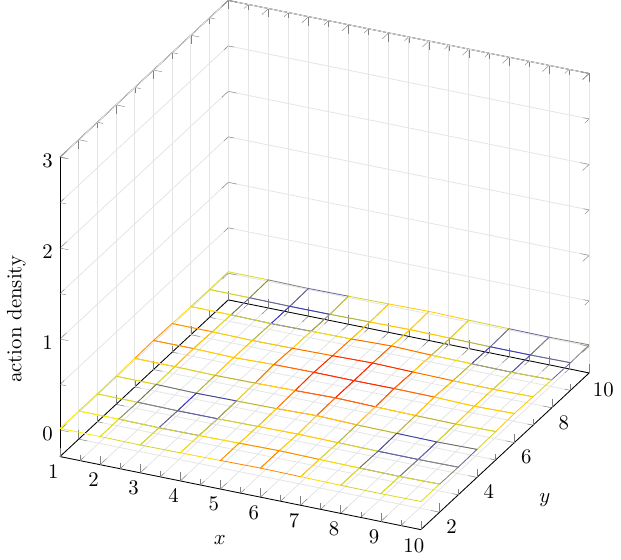}
\caption{$t=5$}
\label{}
\end{subfigure}
\hspace{9mm}
\begin{subfigure}{0.4\columnwidth}
\centering
\includegraphics[width=\columnwidth]{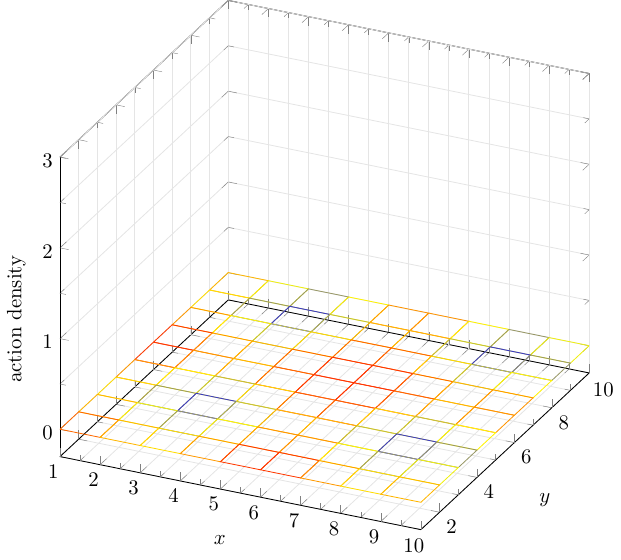}
\caption{$t=8$}
\label{}
\end{subfigure}
\caption{2D distribution of the action density
$-(1/16)\sum_\mu\tr H(x,y,L/2,\mu)^2$ as the function of the flow time~$t$.
$L=10$ and the initial configuration~$g(t=0,x)$ is Eq.~\eqref{eq:(2.4)}
with~$m=1$. The case of the gradient flow with the parameter~$\eta=1$
in~Eq.~\eqref{eq:(2.2)} is shown.}
\label{fig:3}
\end{figure}

\begin{figure}[htbp]
\centering
\begin{subfigure}{0.4\columnwidth}
\centering
\includegraphics[width=\columnwidth]{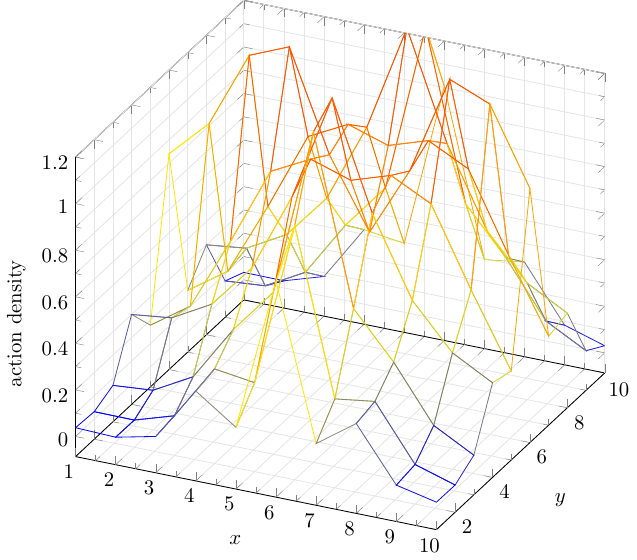}
\caption{$t=0$}
\label{}
\end{subfigure}
\hspace{9mm}
\begin{subfigure}{0.4\columnwidth}
\centering
\includegraphics[width=\columnwidth]{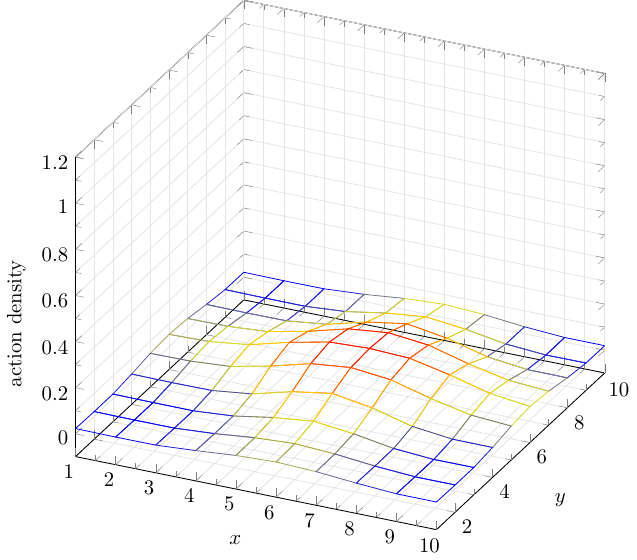}
\caption{$t=1$}
\label{}
\end{subfigure}
\begin{subfigure}{0.4\columnwidth}
\centering
\includegraphics[width=\columnwidth]{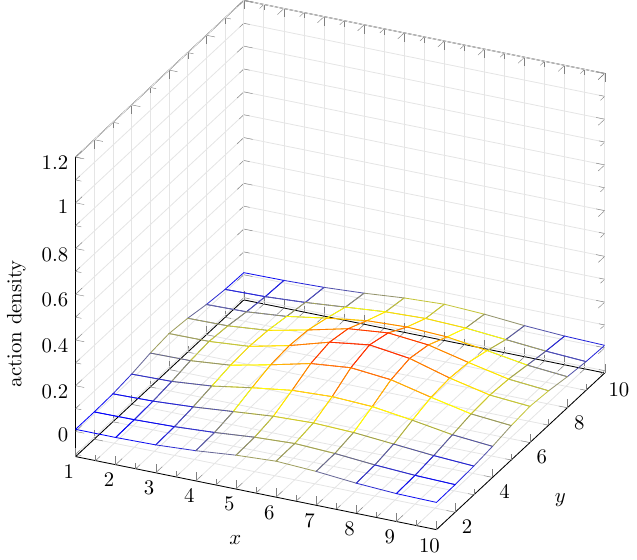}
\caption{$t=8$}
\label{}
\end{subfigure}
\hspace{9mm}
\begin{subfigure}{0.4\columnwidth}
\centering
\includegraphics[width=\columnwidth]{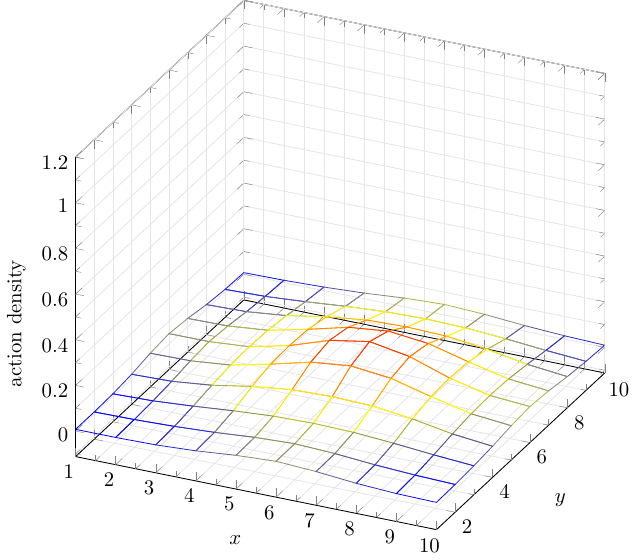}
\caption{$t=15$}
\label{}
\end{subfigure}
\begin{subfigure}{0.4\columnwidth}
\centering
\includegraphics[width=\columnwidth]{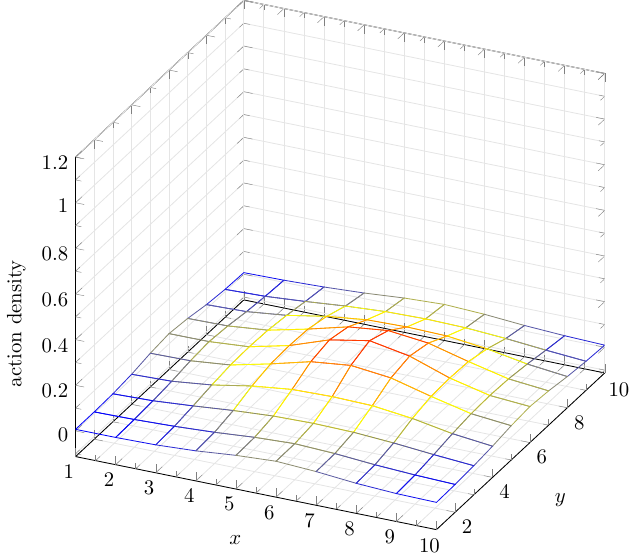}
\caption{$t=25$}
\label{}
\end{subfigure}
\hspace{9mm}
\begin{subfigure}{0.4\columnwidth}
\centering
\includegraphics[width=\columnwidth]{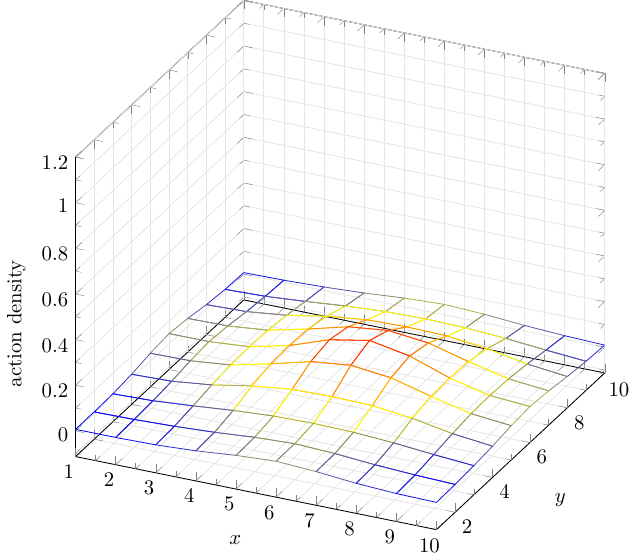}
\caption{$t=40$}
\label{}
\end{subfigure}
\caption{Same as Fig.~\ref{fig:3} but the case of the gradient flow with the
parameter~$\eta=-20$ in~Eq.~\eqref{eq:(2.2)}.}
\label{fig:4}
\end{figure}

We may observe these (de)stabilization effects of the gradient flow as the
evolution of the action density distribution along the flow.
In~Figs.~\ref{fig:3} and~\ref{fig:4}, we see that a non-trivial structure of
the configuration dies out along the flow for~$\eta=1$ (Fig.~\ref{fig:3}) but
it survives for~$\eta=-20$~(Fig.~\ref{fig:4}).

\begin{figure}[htbp]
\centering
\begin{subfigure}{0.45\columnwidth}
\centering
\includegraphics[width=\columnwidth]{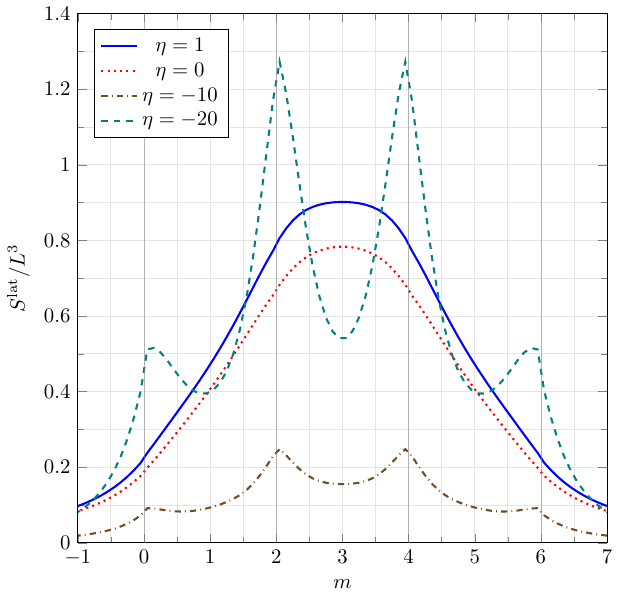}
\caption{$L=10$}
\label{}
\end{subfigure}
\hspace{8mm}
\begin{subfigure}{0.45\columnwidth}
\centering
\includegraphics[width=\columnwidth]{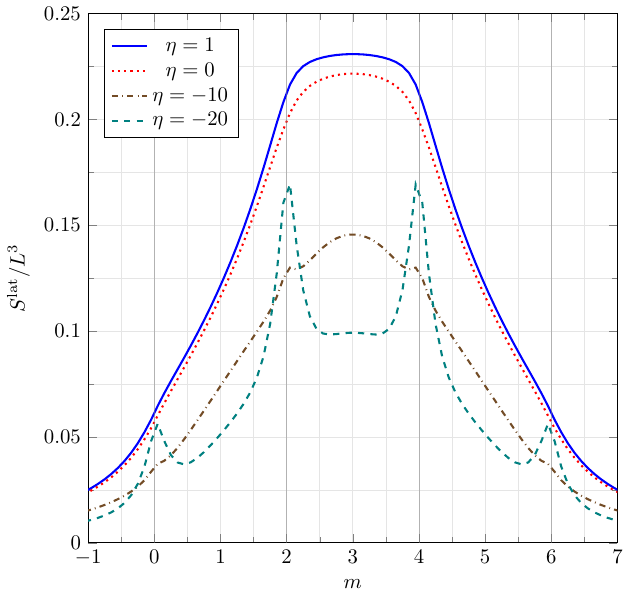}
\caption{$L=20$}
\label{}
\end{subfigure}
\\[5mm]
\begin{subfigure}{0.45\columnwidth}
\centering
\includegraphics[width=\columnwidth]{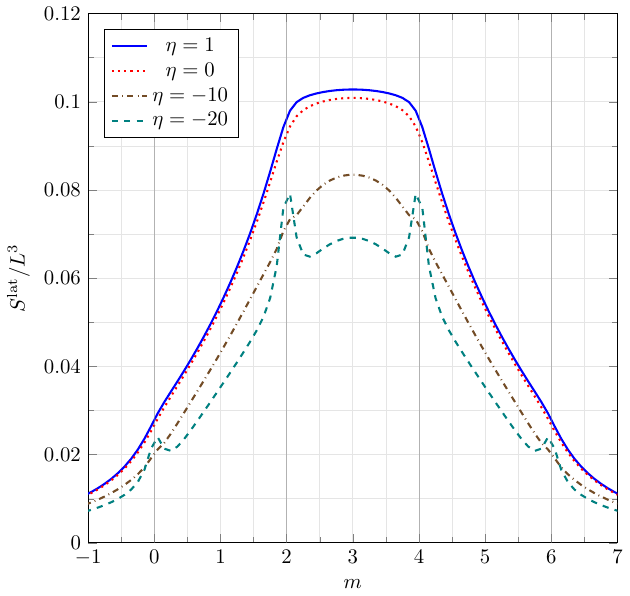}
\caption{$L=30$}
\label{}
\end{subfigure}
\caption{The value of the lattice action~\eqref{eq:(3.7)} of the
mapping~\eqref{eq:(2.4)} as a function of the parameter~$m$ for various
values of~$\eta$. The cases of lattice sizes, $L=10$, $L=20$, and $L=30$, are
depicted.}
\label{fig:5}
\end{figure}

These (de)stabilization of the topological sector under the gradient flow
could be understood by considering the lattice
action~$S^{\text{lat}}$~\eqref{eq:(3.7)} of the mapping~\eqref{eq:(2.4)} as the
function of the parameter~$m$ (this argument is analogous to the study of the
one instanton action as the function of the size moduli
in~Refs.~\cite{Tanizaki:2024zsu,Iwasaki:1983bv,GarciaPerez:1993lic}).
In~Fig.~\ref{fig:5}, we plot the value of the lattice action~$S^{\text{lat}}$ as
the function of~$m$. First, for $\eta=0$ and~$\eta=1$, we see that the lattice
action as the function of~$m$ does not have any local minimum. Since the
lattice action monotonically decreases along the flow
(recall~Eq.~\eqref{eq:(3.6)}), this fact indicates that, for $\eta=0$
and~$\eta=1$, the flow would effectively make the parameter $|m|\to\infty$ that
results in~$g(x)\to\bm{1}$ and the trivial winding number~$W_3^{\text{lat}}=0$.
This well explains the observation which can be made in~Figs.~\ref{fig:1a},
\ref{fig:1b}, \ref{fig:2a}, \ref{fig:2b}, and~\ref{fig:3}. On the other hand,
for~$\eta=-20$ (i.e., the over-improved one), the lattice action as the
function of~$m$ acquires local minima in regions of~$m$ which correspond to
distinct topological sectors. This fact indicates that the lattice winging
number is stable under the gradient flow for~$\eta=-20$ at least until the
lattice size~$L=30$.\footnote{This does not necessarily imply that the flow
preserves the winding number, because here we are assuming a simple form of
the configuration, Eq.~\eqref{eq:(2.4)}. In principle, the configuration could
slip away from the form~\eqref{eq:(2.4)} under the flow, although
Figs.~\ref{fig:1d} and~\ref{fig:2d} indicate that this is not the case.} This
also well explains the behaviors in~Figs.~\ref{fig:1d}, \ref{fig:2d},
and~\ref{fig:4}. We can understand the stabilization effect of the gradient
flow associated with an over-improved lattice action in this way. Note that the
stabilization comes from the discretization error in the flow equation and we
thus expect that the effect becomes weaker for $L$ larger (i.e., when the
lattice becomes finer). However, when the lattice is fine enough, we may simply
use the discretized winding number without relying on the gradient flow; when
the lattice is coarse, we may employ the gradient flow which could provide an
enough stabilization of the winding number. In this way, both elements in our
method can be complementary.

\begin{figure}[htbp]
\centering
\begin{subfigure}{0.45\columnwidth}
\centering
\includegraphics[width=\columnwidth]{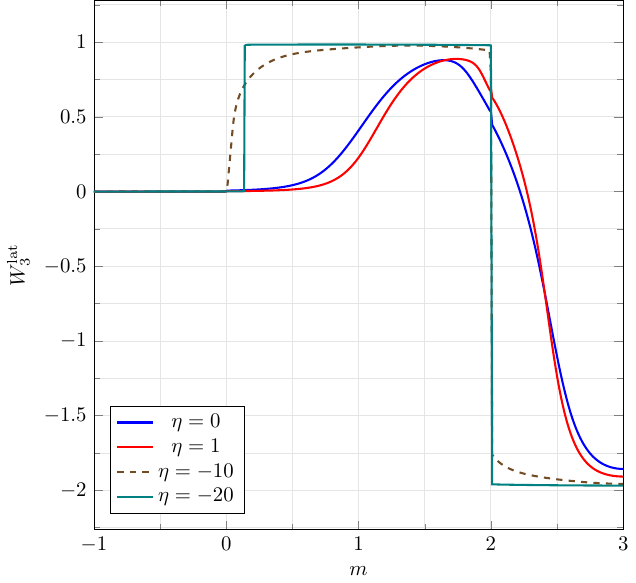}
\caption{$t=2$}
\label{}
\end{subfigure}
\hspace{8mm}
\begin{subfigure}{0.45\columnwidth}
\centering
\includegraphics[width=\columnwidth]{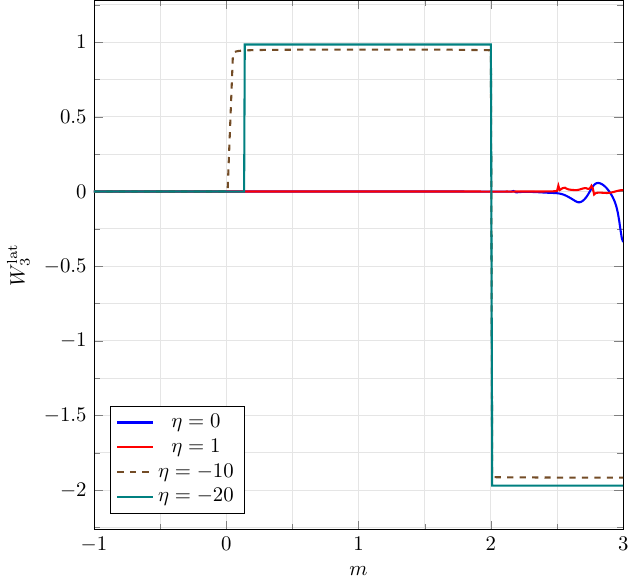}
\caption{$t=10$}
\label{}
\end{subfigure}
\caption{$W_3^{\text{lat}}$ obtained by the gradient flow as the function of the
parameter~$m$. The lattice size is~$L=10$.}
\label{fig:6}
\end{figure}

In Fig.~\ref{fig:6}, we depict $W_3^{\text{lat}}$ obtained by the gradient flow
as the function of~$m$. As Fig.~\ref{fig:5} indicates for, say, $\eta=-20$, the
$W_3^{\text{lat}}$ reproduces the exact values~\eqref{eq:(2.6)} quite impressively
well even for~$L=10$.




\section{Conclusion}
\label{sec:4}
In this paper, we proposed a simple and versatile numerical method which
computes an approximate winding number of a mapping from 3D torus~$T^3$
to~$U(N)$, when $T^3$ is approximated by a discrete cubic lattice. Our method
consists of a tree-level improved discretization of the winding number and the
gradient flow associated with an over-improved lattice action. By employing
a one-parameter family of mappings from $T^3$ to~$SU(2)$ with known winding
numbers, we demonstrated that the method works quite well even for coarse
lattices, reproducing integer winding numbers in a good accuracy. Since we
examined the validity of our method only for a simple one-parameter mappings
with known winding numbers, the implementation of our method for real problems
would require some exploratory analyses especially on the optimal value
of~$\eta$ in the gradient flow which stabilizes the lattice winding number.
Nevertheless, we believe that our method will be practically useful in many
problems because the computationally burden in our method is rather light. We
also note that our method can trivially be generalized to the case of
higher-dimensional tori. Finally, it would also be interesting to generalize
the present method to a 3D lattice with boundary since the resulting system
(approximately) realizes the Wess--Zumino--Witten term~\cite{Witten:1983ar}.

\section*{Acknowledgments}
We would like to thank Soma Onoda and Ken Shiozaki for helpful discussions.
O.M.\ thanks the authors of Ref.~\cite{Tanizaki:2024zsu} for helpful
discussions.
This work was partially supported by Japan Society for the Promotion of Science
(JSPS) Grant-in-Aid for Scientific Research Grant Number JP23K03418 (H.S.).
O.M.\ acknowledges the RIKEN Special Postdoctoral Researcher Program.



%



\let\doi\relax









\end{document}